\documentclass{he_symp}
\usepackage{psfig,graphicx,epsfig}
\usepackage{color}
\usepackage{amsmath,amssymb,epic,eepic,array}
\begin{document}
\renewcommand{\FirstPageOfPaper }{ 100}\renewcommand{\LastPageOfPaper }{ 104}

\title{Chasing the second gamma-ray bright isolated neutron star: 
3EG J1835+5918/RX J1836.2+5925}
\titlerunning{Chasing the 2nd gamma-ray bright INS}
\author{O.Reimer\inst{1}, K.T.S.Brazier\inst{2}, A.Carrami\~nana\inst{3}
G.Kanbach\inst{4}, P.L.Nolan\inst{5}, \and D.J.Thompson\inst{6}}  
\authorrunning{O. Reimer et al.}

\institute{Ruhr-Universit\"at, Theoretische Physik IV, Bochum, Germany,
\and University of Durham, Durham, DH1 3LE, England,
\and INAOE, 72000 Puebla, Mexico,
\and MPE, D-85740 Garching, Germany, 
\and Stanford University, Stanford, CA 94395, USA,
\and NASA GSFC, Greenbelt, MD 20771, USA}
\maketitle

\begin{abstract}
The EGRET telescope aboard NASAs Compton GRO has repeatedly detected 
3EG~J1835+5918, a bright and steady source of high-energy gamma-ray 
emission with no identification suggested until recently. The long
absence 
of any likely counterpart for a bright gamma-ray source located
25$^\circ$ off 
the Galactic plane initiated several attempts of deep observations at
other
wavelengths. We report on counterparts in X-rays on a basis of a 60 ksec 
ROSAT HRI image. In order to conclude on the plausibility of the X-ray 
counterparts, we reanalyzed data from EGRET at energies above 100 MeV 
and above 1 GeV, including data up to CGRO observation cycle 7. 
The gamma-ray source location represents the latest and probably the
final positional assessment based on EGRET data. The X-ray counterparts 
were studied during follow-up optical identification campaigns, leaving
only one object to be likely associated with the gamma-ray source
3EG~J1835+5918.
This object, RX~J1836.2+5925, has the characteristics of an isolated
neutron 
star and possibly of a radio-quiet pulsar.
\end{abstract}

\section{Gamma-Ray Observations}

3EG J1835+5918 was first discovered at photon energies above 100 MeV by
the EGRET instrument aboard NASAs Compton GRO during regularly scheduled 
observations in 1991. The source was repeatedly seen whenever it was in
the 
field of view of the EGRET instrument. However, the first observation
performed 
with a close on-axis pointing towards 3EG~J1835+5918 using EGRET was
only 
made in CGRO observation cycle 7 in 1998. EGRET observations performed
at large 
off-axis viewing angle are problematic due to the degradation of the 
instrumental point spread function (PSF). The first report on GRO
J1837+59 
(\cite{nol94}) included only data from EGRET observations between 1991 
and 1993. The source has been subsequently listed in the EGRET catalogs
as
GRO~J1837+59 (\cite{fic94}) and 2EG~J1835+5919 (\cite{tho95}).
With the appearance of the Third EGRET catalogue (E~$>100$~MeV)
(\cite{har99}) 
and GeV source compilations (E~$>1$~GeV) (\cite{lam97}, \cite{rei97}), 
results from a total of 12 individual observations of 3EG~J1835+5919
were 
reported. This source remained the brightest unidentified EGRET source 
outside the Galactic plane. 

In order to extend the coverage of 3EG~J1835+5919 to its maximum, we
finally  
used all gamma-ray data taken by EGRET through the CGRO mission,
including 
the observations made at a small off-axis angle in observation cycle 7. 
Generally one needs to distinguish between observations in which the
angle 
between 3EG~J1835+5918 and the instrument pointing direction was within
or 
without 25$^\circ$. 
This distinction has been recommended by the EGRET instrument team for 
using the standard PSF (sources within 25$^\circ$ of the instrumental
pointing) 
or using the wide-angle PSF if outside. 
The EGRET observations from CGRO observation cycle 7 extend
significantly 
beyond the catalogued observations. They are separated by more than 3
years 
from the previous observations of 3EG~J1835+5918. Both the
long-term observational aspect and the quality of the observation have
been improved: despite the lower efficiency of the EGRET spark
chamber, the 1998 observations were the first on-axis observations,
unbiased 
from effects which most of the earlier EGRET observations of
3EG~J1835+5918 
could suffer from. The narrow field-of-view mode, used in EGRET viewing
periods 710 and 711, does not introduce additional problems in this
respect.

The best source location was obtained from EGRETs highest energetic
photons, 
where the instrumental PSF is significantly smaller than at lower
energies. 
Using a likelihood method (\cite{mat96}), \cite{rei01} determined 
the best position ($> 1$ GeV) to be l = 88.80$^\circ$, b =
25.02$^\circ$, 
which is consistent with the position from an analysis above 100 MeV 
(l = 88.76$^\circ$, b = 25.09$^\circ$), the position given in 3EG and 
GEV catalogues, and the elliptical fit from \cite{mat01}
(l = 88.74$^\circ$, b = 25.08$^\circ$, 
a = 9.7$\arcmin$, b = 7.8$\arcmin$, $\Phi$ = 13$^\circ$). 
However, the additional data gave positional errors of only 6$\arcmin$
and 8$\arcmin$ 
for the $68 \%$ and $95 \%$ confidence region, respectively. 

Nolan et al. 1994 and \cite{mcl96} indicated flux variability for
3EG~J1835+5918 
on the basis of smaller data sets than presented lastly. 
The most recent variability study (\cite{tom99}) 
puts 3EG~J1835+5918 clearly among the non-variable sources, similar to 
the identified gamma-ray pulsars. Tompkins made use of an algorithm 
especially adopted for the characteristics of the observations by EGRET, 
i.e. sparse data sets from individual observations, often widely
separated 
in time and characterized by different background levels. In addition, 
data from individual EGRET observations up to CGRO observation 
cycle 4 were used in this study. A strict data selection 
(only within 25$^\circ$ on-axis) among the gamma-ray observations was
used, 
assuring a data set of comparable quality.

We complemented the flux history of 3EG~J1835+5918 with the data 
from 13-27 January 1998, the last high-energy gamma-ray data on 
this source to be taken for some years. Due to the generally lower 
efficiency of the EGRET spark chamber towards the end 
of the EGRET mission, the early 1998 viewing periods were evaluated
using adjusted 
normalization factors (\cite{esp99}). These factors were checked 
quantitatively by means of a similar on-axis observation of Geminga
during 
7-21 July 1998. Assuming that the instrumental sensitivity has not
changed 
appreciably between these observations and that Geminga remains the
stable 
gamma-ray emitter previously observed, this normalization for Geminga
could 
be applied to the flux of 3EG~J1835+5918 in the cycle 7 observations.
In addition, one needs to consider observations with up to 25$^\circ$
off-axis 
separately from observations outside 25$^\circ$. The fluxes above 100
MeV and above 
1 GeV appear to be linearly correlated, considering the uncertainties in
individual 
viewing periods arising from photon statistics, especially for the
sparse data of 
the detections above 1 GeV. 

We concluded, there is no indication of flux variability after all for
3EG~J1835+1918, 
neither above 100 MeV nor above 1 GeV. Given the differing quality of
the EGRET 
observations within their statistical and systematical uncertainties, we 
find 3EG J1835+5918 compatible with a non-variable source of an average
flux
of 5.9 $\times$ 10$^{-7}$ cm$^{-2}$ s$^{-1}$ (E $> 100$ MeV).

After realizing that 3EG~J1835+5918 is consistent with having constant 
gamma-ray flux throughout the EGRET mission, the issue of its spectral 
variability still remains. Nolan et al. 1996 reported evidence for 
spectral variability between individual EGRET viewing periods.
Apparently, 
no correlation between spectral index and flux was found.
Hence, we re-examined the EGRET data on 3EG~J1835+5918 for indication 
of spectral variability. Individual spectra in each of the relevant
viewing 
periods were determined by simultaneously analyzing likelihood excesses
of 
3 $\sigma$ detection significance and above. We derived a flux value or
upper 
limit in each of ten energy intervals (30 MeV to 10 GeV) using a 
likelihood method. In cases when poor count statistics gave a spectrum
dominated
by upper limits, the ten energy intervals were recombined into four
(30-100, 100-300, 300-1000, $>$1000 MeV), followed by the appropriate 
determination of the spectral slope. Also, when the source position
determined from likelihood analysis of an individual observation 
differed from the GeV-position, both positions were individually 
considered for consequences for the resulting spectrum. None of them
introduces
relevant modifications in the resulting spectral slope. Therefore, the 
determined individual spectra could be compared at the best level
currently
achievable for an unidentified high-energy gamma-ray source.

We find that the spectra of 3EG~J1835+5918 determined from individual 
viewing periods are fully compatible within their statistical and
systematic 
uncertainties throughout the entire EGRET mission. A single power law
spectral 
index of 1.73 $\pm$ 0.07 is consistent within 1 $\sigma$ for all
individual spectra.

With the consistency of the individual spectra throughout the EGRET
observations established, we co-added the data from cycles 1 to 7 in
order to 
determine the best overall spectrum of 3EG~J1835+5918. A single
power-law fit appears to 
be inadequate for this source. The spectrum of 3EG~J1835+5918 resembles
the gamma-ray 
spectra of known gamma-ray pulsars like Vela or Geminga and the spectra
of 
candidate gamma-ray pulsars 3EG J2020+4017, 3EG J0010+7309, and 3EG
J2227+6122: 
the hard power law spectral index, as determined to be -1.7 $\pm$ 0.06
between 70 MeV and 4 GeV, 
the high-energy spectral cut-off or turnover as well as a possible
spectral softening at 
the low energies. 
Upper limits from COMPTEL do not constrain the shape of the spectrum
at lower energies. The TeV upper limits as reported by Whipple are
consistent
with a rollover at 4 GeV, but certainly not with a simple extrapolation
of the 
EGRET measured power law spectrum to even higher energies. 

\section{Radio Observations}

Deep searches in radio (770 MHz) at the position of 2EG J1835+59 could
not 
detect any object above 0.5 Jy (\cite{nic97}). This is in agreement with 
the correlation study between unidentified EGRET sources and catalogued 
flat-spectrum radio sources (Green Bank 4.85 GHz/Parkes-MIT-NRAO 4.85
GHz),
not suggesting any radio counterpart for 3EG J1835+59 (\cite{mat01}). 

\section{X-Ray Observations}

With the analysis of the 60 ksec ROSAT High Resolution Imager
observation 
from December 1997/January 1998, the earlier coverage of this source in 
X-rays could be increased by a factor of 12. For the first time,
counterparts 
have been discovered in X-rays between 0.1 -2.4 keV (\cite{rei99}).
The sources are all faint with HRI count rates of 1-3 ksec$^{-1}$. 
Only sources 1 and 10 are excluded due to positional disagreement with
the EGRET source,
see Fig. 1 

\begin{figure}
\centerline{\psfig{file=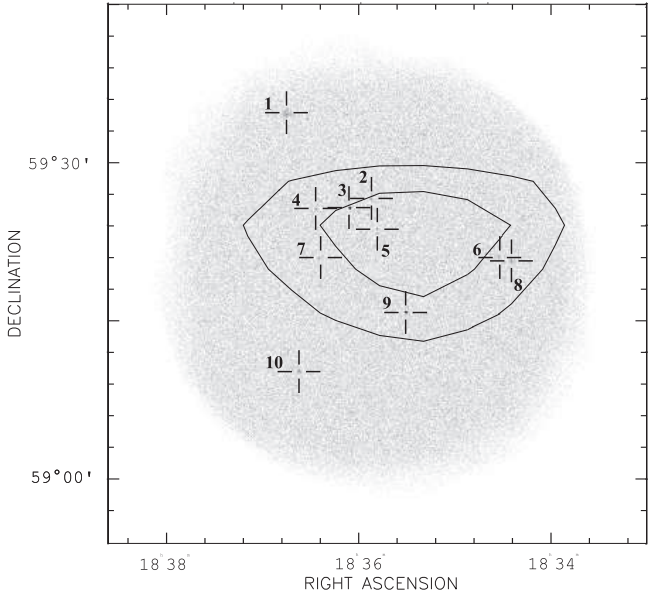,width=8.8cm,clip=} }
\caption{The long ROSAT HRI (0.1 - 2.4 keV) observation of the 
field of 3EG J1835+5918 from December 1997/January 1998. The X-ray image
is overlaid 
with source location contours ($68\%$ and 95$\%$) of the high-energy
gamma-ray source, 
determined above 1 GeV. The detected X-ray sources are indicated and
were subject of a 
optical follow-on identification campaign.
\label{image}}
\end{figure}

\section{Optical Observations}

The discovered X-ray sources were subject of deep optical studies, 
performed at the 2.1 m telescope at the Guillermo Haro observatory at 
Cananea, Mexico. The instrumentation is aimed for ROSAT counterpart 
identifications and served well for our purpose. The observations were 
accompanied by studies of the DSS-2 images and use of the USNO A2.0
catalog. 
Independently, Halpern \& Mirabal studied these counterparts at 
the 2.4 m Hilter telescope, the 3 m Shane reflector at Lick and 
the Hobby-Eberle telescope. 
The following optical counterparts were found:\\
1)  positional disagreement, QSO with z = 0.46\\
2)  QSO, z = 1.87\\
3)  U $>$ 22.3,B $>$ 23.4,V $>$ 25.2,R $>$ 24.5\\
4)  V = 18.9, R = 19.2, QSO with z = 1.75\\
5)  V = 19.3, R =19.1, QSO with z = 1.865\\
6)  M5V star\\
7)  QSO with z = 1.36\\
8)  G7V dwarf star\\ 
9)  bright object in center (V = 15.6), M type star, 
 fainter object out of X-ray source error box, G type dwarf\\
10)  K5V-star, in positional disagreement\\

\section{Conclusions}

3EG~J1835+5918 is a persistent high-energy gamma-ray source located at  
high Galactic latitudes and has been observed repeatedly by EGRET. It is 
characterized by a hard power law and a spectral break or turn-over
above 4 GeV. 
It appears to be a non-variable source in terms of its flux as well as 
its spectral shape throughout the entire EGRET mission, despite
suggestions of 
variability from earlier analysis. Its gamma-ray properties are typical
of those
observed from other gamma-ray pulsars and candidate radio-quiet neutron
stars.
The deep ROSAT HRI observation revealed several X-ray sources consistent
with 
the location of the observed GeV-emission of 3EG~J1835+5918. As a result
of the 
identification campaigns independently carried out by \cite{mir00} and
ourselves, 
only one of the ten X-ray sources still attracts interest to be
considered further for 
an association with the $\gamma$-ray source. This source,
RX~J1836.2+5925,
is characterized by an obvious lack of radio-emission, indetectibility
by means
of an UV-excess identification technique, lack of optical counterpart up
to
V$\sim$25 mag, and location well inside the 68\% likelihood test 
statistic contours of 3EG~J1835+5918. Our HRI observation contain no
information on the 
X-ray spectrum of RX~J1836.2+5925. Hence, assuming that this X-ray
source is the most 
likely counterpart to 3EG J1835+5918, we are restricted to using the
X-ray flux of 
RX~J1836.2+5925 and the gamma-ray properties of 3EG~J1835+5918
to investigate the characteristics of the object.

To do so, we can use its multi-frequency properties to ascertain its
characteristics. 
The high $F_\gamma$/$F_{radio}$ value seems to rule out a blazar origin.
The already 
noted similarities in the gamma-ray characteristics with known gamma-ray
pulsars 
or radio-quiet pulsar candidates preferably suggests a neutron star
nature for 
3EG~J1835+5918/RX J1836.2+5925. 
We have reexamined the gamma-ray and X-ray fluxes for all the known and
candidate 
pulsars, using a consistent energy range in both bands.  
Comparing the flux of 3EG~J1835+5918 in $\gamma$-rays (E $> 100$ MeV)
and RX J1836.2+5925 in X-rays (0.12 - 2.4 keV), this source falls among the 
candidates currently considered for associations between gamma-ray
sources  and X-ray sources with proven or suspected neutron star origin, 
see Fig.5. Nearly all of the candidate gamma-ray pulsars lie at the bottom 
end of
the  sensitivity feasible for the last generation of X-ray instruments like
ROSAT, 
ASCA, and SAX. Obviously, only deep observations could reveal
counterparts at 
all or with features not easily explained by any other astronomical
objects. 
The lack of optical counterparts up to V$\sim$25 mag and radio emission
for 
RX~J1836.2+5925 is a further characteristic signature for isolated,
radio-quiet 
neutron stars, and ideally demonstrated by Geminga as its gamma-ray
bright 
prototype.
 
Although many of the candidate radio-quiet pulsars beside Geminga itself
are 
located within or near SNRs, 3EG~J1835+5918 does not. Neither radio
observations
nor the X-ray data yield any hint of a SNR in the vicinity of this
object, and
the high Galactic latitude seems to rule out the possibility of
obscuration that 
might hide one.

If 3EG~J1835+5918/RX~J1836.2+5925 is not of quasar origin and also not
the 
first candidate of an hypothesized extragalactic astronomical object
bright 
and steady in gamma-rays, faint in X-rays, and yet undetectable at
optical 
and radio wavelengths, it will reside within our Galaxy. We therefore
have 
to suspect an isolated radio-quiet neutron star candidate. With Geminga
as 
the only established pulsar of a predicted class of radio-quiet pulsars,
a 
comparison of observational parameters in analogy with 
3EG J1835+5918/RX J1836.2+5925 might be appropriate. 
Fig 2 compares the multi-frequency $\nu$F$\nu$-spectrum of Geminga and
3EG~J1835+5918, 
respectively.

\begin{figure}
\centerline{\psfig{file=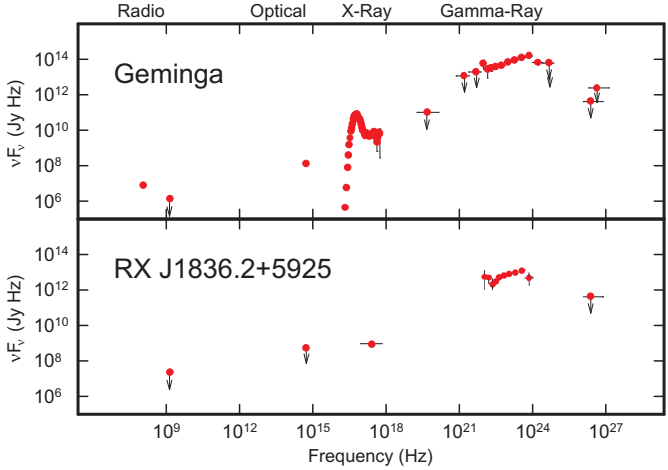,width=8.8cm,clip=} }
\caption{Multi-frequency energy spectra of Geminga 
and RX J1836.2+5925/3EG J1835+5918. 
Similarities could easily be seen, however the spectral 
energy distribution is not nearly as good covered in 
the case of RX J1836.2+5925. Completing the SED 
in X-rays between 0.1 and 10 keV or revealing an 
optical counterpart will be the most promising way to 
confirm RX J1836.2+5925/3EG J1835+5918 as the 
2nd gamma-ray bright radio-quiet isolated neutron star.
\label{image}}
\end{figure}

\begin{figure}
\centerline{\psfig{file=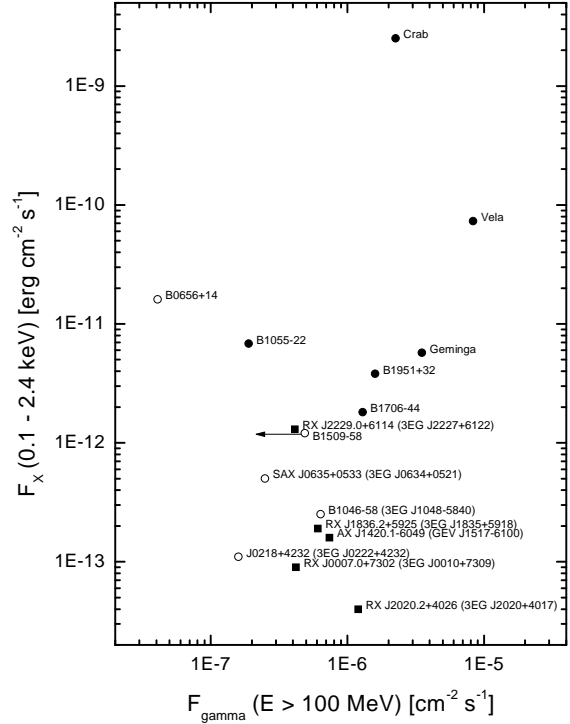,width=8.8cm,clip=} }
\caption{X-ray and gamma-ray fluxes of high-confidence 
pulsar detections (filled circles), probable associations between
pulsars and 
high-energy gamma-ray sources (open circles), and candidate radio-quiet
pulsars
(filled squares).
All X-ray fluxes are given for the 0.1 to 2.4 keV energy band, in cases
of
different energy band quoted in the literature, the flux is normalized 
into the chosen energy band. The gamma-ray fluxes are given above 100
MeV, 
in cases where different event selection criteria were used the
appropriate 
gamma-ray flux above 100 MeV has been determined for the 
energy band desired here. 
\label{image}}
\end{figure}

First, Geminga is three times brighter in gamma-rays and about fifty
times 
brighter in X-rays. 
To extrapolate from the distance to Geminga using the observed fluxes, 
3EG~J1835+5918 would lie between 250 pc (scaling from gamma-rays) and 
1.1 kpc (scaling from X-rays), assuming the same beaming as Geminga. 
Besides, pulsars tend to begin their life in the Galactic plane. 
A pulsar moving with a typical velocity of about 350 km/s would move 
only 300 pc, even in a lifetime of 106 years, while an object seen at b
= 25$^\circ$ 
would have to move more than 420 pc from the plane if it were at a
distance 
greater than 1 kpc. As pointed out by \cite{yad95} discussing 
the beaming evolution of pulsars in the outer-gap model, the beaming
fraction 
becomes rather small as the pulsars age increases. Therefore a distant
but 
old pulsar would have to be immensely powerful or exceptionally beamed. 
If 3EG~J1835+5918 is more distant, then its gamma-ray luminosity would 
exceed that of Geminga, but if closer, then the surface brightness in
X-rays 
of the neutron star would have to be lower than Gemingas. This
indicates, that 
either the efficiency of the emission mechanism is different and/or the 
parameter space which radio-quiet pulsar candidates could occupy is 
wide spread. 
In contrast to energetic pulsars like Vela or B1706-44, non-thermal
emission 
or PWN features have not been observed here so far. Nor is it an
extended 
source in X-rays. The lack of an associated SNR as well as the rare
chance 
to find a similar pulsar at such high Galactic latitude (to say: nearby) 
argues against a young pulsar in the case of 3EG~J1835+5918. 
However, the striking similarities in the gamma-ray properties between 
Geminga, other candidate radio-quiet pulsars and 3EG~J1835+5918, the
absence 
of a radio and optical counterpart of RX~J1836.2+5925, points 
toward its nature of an old but radio-quiet neutron star.

Certainly, neither the X-ray data nor the gamma-ray data currently 
allow wide range period scans for pulsations without known ephemeris
(\cite{jon98}, \cite{cha01}). Any (potentially) successful search for 
periodicity will have to be postponed until more sensitive instruments 
like XMM in the X-rays or GLAST in the gamma-rays will have observed 
3EG~J1835+5918. However, if a restrictive set of parameters can be 
predicted from pulsar models or if a lightcurve can be derived from 
another wavelength, the archival EGRET data will permit the discovery of 
pulsations in the gamma-rays. The long observational history presented 
here will certainly assist in any such effort. 
Finally, RX J1836.2+5925 might be identified as a neutron star by 
extremely deep optical imaging/spectroscopy, as already in progress
using the
Subaru telescope (\cite{kaw02}).\\ 
To unambiguously relate 3EG~J1835+5918 to a known class of astronomical
objects 
would be of extreme importance for any collective studies of gamma-ray
sources,
and a gain for general pulsar physics if the existence of another 
isolated neutron star in gamma-rays will be confirmed.



\clearpage


\begin{thebibliography}{} 

\bibitem[Chandler et al. 2001]{cha01}Chandler, A.M. et al., ApJ 556,
(2001), 59
\bibitem[Esposito et al. 1999]{esp99} Esposito, J.A. et al., 1999, ApJS,
123, 203 
\bibitem[Fichtel et al. 1994]{fic94} Fichtel, C.E. et al., ApJS, 94,
551, 
\bibitem[Hartman et al. 1999]{har99} Hartman, R.C. et al., 1999, ApJS,
123, 79
\bibitem[Jones 1998]{jon98} Jones, B.B., 1998, PhD thesis,
astro-ph/0202088
\bibitem[Kawai et al. 2002]{kaw02} Kawai, N., Totani, T. and Kawasaki,
W., these proceedings
\bibitem[Lamb \& Macomb 1997]{lam97} Lamb, R.C. and Macomb, D.J., 1997,
ApJ, 488, 872
\bibitem[Mattox, Hartman \& Reimer 2001]{mat01} Mattox, J.R. et al.,
2001, ApJS 135, 155
\bibitem[Mattox et al. 1996]{mat96} Mattox, J.R. et al., 1996, ApJ, 461,
369
\bibitem[McLaughlin et al. 1996]{mcl96} McLaughlin, M.A. et al., 1996,
ApJ, 763
\bibitem[Mirabal \& Halpern 2000]{mir00} Mirabal, N. and Halpern, J.,
2000, ApJ 541, 180
\bibitem[Nice \& Sayer 1997]{nic97} Nice, D.J. and Sawyer, R.W., 1997,
ApJ, 476, 261
\bibitem[Nolan et al. 1994]{nol94} Nolan, P.L. et al., in Proc. 2nd
Compton Symposium, 1994, 
AIP Conference Proceedings 304, 360
\bibitem[Nolan et al. 1996]{nol96} Nolan, P.L. et al., 1996, ApJ, 459,
100
\bibitem[Reimer et al. 1997]{rei97} Reimer, O. et al., in Proc. 25th
ICRC, 1997, Vol.3, 97 
\bibitem[Reimer et al. 1999]{rei99} Reimer, O. et al., in Proc. 5th
Compton Symposium, 1999, 
AIP Conference Proceedings 510, 489
\bibitem[Reimer et al. 2001]{rei01} Reimer, O. et al., 2001, MNRAS 324,
772
\bibitem[Thompson et al. 1995]{tho95} Thompson, D.J. et al., 1995, ApJS,
101, 259
\bibitem[Tompkins 1999]{tom99} Tompkins, W., 1999, PhD thesis,
astro-ph/0202141
\bibitem[Yadigaroglu \& Romani 1995]{yad95} Yadigaroglu, I.A. and
Romani, R.W., 1995, ApJ, 449, 211

\end{thebibliography}
\end{document}